\documentclass[10pt,aps,prb,superscriptaddress,reprint]{revtex4-1}
\usepackage{graphicx}
\usepackage{amsmath}
\usepackage[english]{babel}
\usepackage{array}
\usepackage{amssymb, bm}
\usepackage{multirow}
\usepackage{epstopdf}

\begin{document}
	\title{Electromagnetic proximity effect controlled by spin-triplet correlations in superconducting spin-valve structures}
	
	\author{Zh. Devizorova}
	\affiliation{Moscow Institute of Physics and Technology, 141700 Dolgoprudny, Russia}
	\affiliation{Kotelnikov Institute of Radio-engineering and Electronics RAS, 125009 Moscow, Russia}
	\affiliation{Institute for Physics of Microstructures, Russian Academy of Sciences, 603950 Nizhny Novgorod, GSP-105, Russia}
	\author{S. V. Mironov}
	\affiliation{Institute for Physics of Microstructures, Russian Academy of Sciences, 603950 Nizhny Novgorod, GSP-105, Russia}
	\author{A. S. Mel'nikov}
	\affiliation{Institute for Physics of Microstructures, Russian Academy of Sciences, 603950 Nizhny Novgorod, GSP-105, Russia}
	\author{A. Buzdin}
	\affiliation{University Bordeaux, LOMA UMR-CNRS 5798, F-33405 Talence Cedex, France}
	\affiliation{Sechenov First Moscow State Medical University, Moscow, 119991, Russia}

	\begin{abstract}
		The spin-triplet correlations in superconducting spin valve structures arising in the presence of noncollinear textures of magnetic moment
		are shown to enhance strongly the electromagnetic proximity effect, i. e. the long-range leakage of the magnetic field from the ferromagnet (F) to the superconducting (S) layer. Both the dirty and clean limits are studied on the basis of the Usadel and Eilenberger theory, correspondingly. Our results suggest a natural explanation for the puzzling enhancement of the spontaneous magnetic fields induced by the noncollinear magnetic structures observed by the muon spin rotation techniques in a wide class of layered S/F systems. We show that the electromagnetic proximity effect causes the shift of the Fraunhofer dependence of the critical current on the external magnetic field in the Josephson junction with one superconducting electrode covered by the ferromagnetic layer. This provides an alternative way to measure both the magnitude and the direction of the spontaneous magnetic field induced in the superconductor. We also demonstrate the possibility of the long ranged superconductivity control of the magnetic state in F$_1$/S/F$_2$ structures.
	\end{abstract}

	\maketitle
	
	\section{Introduction}
	
	The penetration of the Cooper pairs from the superconductor (S) to ferromagnets (F) and normal metals is widely known as a manifestation of the so called "proximity" effect which results in a number of spectacular phenomena studied intensively during several decades \cite{Buzdin_RMP, Golubov_RMP, Bergeret_RMP}. No wonder that this proximity effect is accompanied by the 
	back-action of the ferromagnet on the superconducting subsystem revealing in the leakage of the magnetic moments and fields through the S/F interface. The resulting magnetic fields induced inside the superconductor have been experimentally observed with a variety of techniques including the nuclear magnetic resonance \cite{Garifullin}, polar Kerr effect analysis \cite{Xia}, neutron scattering \cite{Khaydukov, Nagy, Ovsyannikov}, and muon spin rotation measurements \cite{Di_Bernardo, Lee_NatPhys, Lee_PRL}.
	
	It is commonly believed that in S/F systems there are only two major mechanisms responsible for the leakage of the magnetic field from the F to the S layer. The first one is associated simply with the stray magnetic field which penetrates into the superconductor and induce the screening currents there (orbital effect) \cite{Aladyshkin}. The second one (so-called inverse proximity effect) is attributed to the the spin polarization of electrons forming the Cooper pair arising near the S/F interface \cite{Krivoruchko, Bergeret_IPE, Bergeret_IPE_Clean, Lofwander, Faure}. Indeed, the electron with the spin along the exchange field easily penetrates the F-layer while the electron with the opposite spin tends to stay in the superconductor. As a result, the opposite electron spins appear to be spatially separated which gives rise to the spin polarization and subsequent magnetization of the superconducting surface layer with the width of the order of the Cooper pair size, i.e., the superconducting coherence length $\xi\sim 1-100 ~{\rm nm}$. 
	
	However, the recent experiments on muon and neutron scattering have revealed the anomalously large distances of the magnetic field penetration to the superconductor in V/Fe, Au/Nb/ferromagnet, Cu/Nb/Co and YBaCuO/LaCaMnO structures \cite{Khaydukov, Lee_NatPhys, Flokstra_PRL, Stahn}. These distances exceed the corresponding values of $\xi$ up to five times which is inconsistent with the predictions of the inverse proximity effect theory. At the same time, the in-plane orientation of magnetic moment in the F layers and the absence of the magnetic domains rule out the orbital effect. 
	
	An alternative explanation of the long-range magnetism in planar S/F structures is based on the  electromagnetic proximity effect\cite{Mironov_APL_2018}. This effect originates from the generation of the superconducting currents {\it inside} the F layer due to the direct proximity effect and the subsequent appearance of the compensating Meissner currents flowing in the S layer. The magnetic field induced by the Meissner currents decays at the distances of the order of the London penetration depth $\lambda$ which can naturally explain the observed long-range magnetism in type-II superconductors where $\lambda > \xi$. Some hints about the screening supercurrents flowing outside the ferromagnet in thin S/F structures have been also obtained in Refs.[\onlinecite{Bergeret_EPL},\onlinecite{Krawiec_PRB}].
	
	\begin{figure}[t!]
		\includegraphics[width=0.35\textwidth]{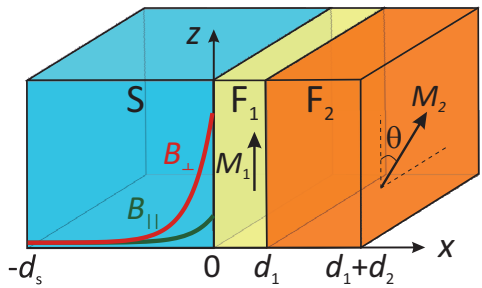}
		\caption{The sketch of the S/F$_1$/F$_2$ system where the non-collinearity of the magnetic moments strongly enhances the magnitude of the electromagnetic proximity effect. } \label{Fig_System}
	\end{figure}

	Interestingly, the muon-spin rotation measurements performed for the Au/Nb/ferromagnet structures indicate the puzzling behavior of the induced magnetic field in the case when the F layer consists of two ferromagnets with different orientations of magnetic moments in the plane of the sample \cite{Lee_NatPhys}. One can naively expect that the magnetic field induced inside the S layer should be stronger for the parallel orientation of magnetic moments in comparison with the perpendicular one since the average exchange field (or magnetization) in the structure, which is the ``source'' of this field, is larger in the former case. However, the experimental data show that the situation is opposite: the observed magnetic field is more pronounced if the magnetic configuration of the ferromagnet is non-collinear.
	
	In the present paper we develop the theory of the electromagnetic proximity effect in the S/F$_1$/F$_2$ structures (see Fig.~\ref{Fig_System}).  Contrary to the previous work Ref.[\onlinecite{Mironov_APL_2018}] where S/F bilayer was considered here we concentrate on the structures with composite F layer consisting of two ferromagnets with different orientations of magnetizations. In spite of some qualitative results about the behavior of the magnetic field induced in the superconductor in such structures was also presented in Ref.[\onlinecite{Mironov_APL_2018}] the detailed microscopical description was lacking. Here we calculate this magnetic field both in dirty and clean limits and  demonstrate that it is much stronger for the perpendicular orientation of the magnetic moments in two ferromagnets compared to the parallel one. Such anomalous behavior of the induced field is attributed to the  appearance of additional equal-spin triplet correlations in the case of non-collinear magnetic moments. Since these correlations are long-ranged they give rise to the significant enhancement of the supercurrent flowing in the F layer, thus, increasing the amplitude of the magnetic field induced in the superconductor. Our results provide a natural explanation of the experimental data reported in Ref.[\onlinecite{Lee_NatPhys}]. We propose an alternative way for experimental determination of both the modulus and the direction of the magnetic field induced in the superconductor. It is based on the fact that the electromagnetic proximity effect causes the shift of the Fraunhofer dependence of the critical current on the external magnetic field in the Josephson junction where one electrode is covered by the ferromagnet. We also demonstrate that in F$_1$/S/F$_2$ structures due the electromagnetic proximity effect antiparallel magnetic configuration is more favorable.
	
	The paper is organized as follows. In Sec. II we consider S/F$_1$/F$_2$ structure and calculate the magnetic field induced in the superconductor for parallel  and perpendicular orientations of the magnetic moments in the ferromagnets in dirty and clean limits. In Sec. III we calculate the Fraunhofer dependence of the critical current vs. the external magnetic field in S$_1$/I/S$_2$/F Josephson junction. In Sec. IV we analyze the influence of the electromagnetic proximity effect on magnetic configuration in F$_1$/S/F$_2$ type structures. In Sec.V we summarize our results.

	\section{Electromagnetic proximity effect in S/F1/F2 structures}
	
	We consider the S/F$_1$/F$_2$ structure consisting of the superconductor (S) of the thickness $d_s \gg \lambda$ and two ferromagnets F$_1$ and F$_2$ with the thicknesses $d_1 \ll \lambda$ and $d_2 \ll \lambda$, respectively (see Fig.\ref{Fig_System}). In the F$_1$ layer the magnetization ${\bf M}_1=M_0 {\bf e}_z$ is directed along the $z$ axis, while the magnetization ${\bf M}_2$ of the F$_2$ ferromagnet forms the angle $\theta$ with the $z$ axis, so that ${\bf M}_2=M_0 \sin \theta {\bf e}_y + M_0 \cos \theta {\bf e}_z$. The $x$ axis is chosen to be perpendicular to the layers with $x=0$ in the S/F$_1$ interface.
	
	To describe the magnetic field arising in the system we choose the vector potential ${\bf A}$ to have only two components ${\bf A}=A_y(x){\bf e}_y + A_z(x) {\bf e}_z$. According to the Maxwell theory, this vector potential satisfies the equation
	\begin{equation}
	\label{MaxEq}
	{\rm rot}~ {\rm rot} {\bf A}=\frac{4\pi}{c}({\bf j}_s+{\bf j}_m)
	\end{equation}
	where ${\bf j}_s$ is the superconducting current and ${\bf j}_m=c~ {\rm rot} {\bf M}$ is the magnetization current flowing at the boundaries of the ferromagnetic layers. 
	
	Our strategy is to use Usadel and Eilenberger approaches to calculate the material relation ${\bf j}_s({\bf A})$ for the dirty and clean limits, respectively, and then use this relation to solve Eq.~(\ref{MaxEq}) and obtain the dependence of the magnetic field induced in the superconductor on the angle $\theta$ between the magnetic moments. 
	
	\subsection{Dirty limit}\label{Sec_Dirty}
	
	In the dirty limit the relation between the supercurrent and the vector potential is local:
	\begin{equation}
	\label{LEq}
	{\bf j}_s(x)=-\frac{c}{4\pi \lambda^2(x)}{\bf A},
	\end{equation}
	where $\lambda(x)$ is the London penetration depth. In the superconductor far away from the S/F$_1$ interface the density of the superconducting electrons is uniform and, consequently, the length $\lambda$ does not depend on $x$ (in what follows we will denote the London penetration depth in the bulk of the superconductor as $\lambda_0$). However, in the small region of the thickness $\xi$ near the S/F$_1$ interface the penetration of the Cooper pairs into the ferromagnet results in the renormalization of $\lambda$. The contribution to the final result coming from such renormalization has the order of $\xi/\lambda\ll 1$ and can be neglected \cite{Mironov_APL_2018}. Then the solution of Eq.~(\ref{MaxEq}) inside the S layer gives  $ {\bf A}(x)={\bf A}_0\exp(x/\lambda_0)$ where ${\bf A}_0$ is the vector potential at the S/F$_1$ interface. The corresponding magnetic field reads $ {\bf B}(x)={\bf B}(0)\exp(x/\lambda_0)$, where $B_y(0)=-A_{0z}/\lambda_0$ and $B_z(0)=A_{0y}/\lambda_0$.
	
	To solve Eq.~(\ref{MaxEq}) inside the ferromagnets we may neglect the effect of the Meissner currents on the spatial variation of the vector potential since the thicknesses of the F$_1$ and F$_2$ layers are assumed to be much smaller than $\lambda$. Then the vector potential inside the ferromagnets takes the following form: ${\bf A}=[A_{0y}+4\pi M_0 x] {\bf e}_y + A_{0z}{\bf e}_z$ for $0<x<d_1$ and ${\bf A}=[A_{0y}+4\pi M_0 d_1 +4\pi M_0(x-d_1)\cos \theta]{\bf e}_y + [A_{0z}-4\pi M_0(x-d_1)\sin \theta]{\bf e}_z$ for $d_1<x<d_2$.
	
	To find the constants $B_y(0)$ and $B_z(0)$ we substitute the above expressions for the vector-potential into the equation (\ref{MaxEq}) and integrate it over the width of the ferromagnets $d=d_1+d_2$ taking into account that ${\bf B}={\rm rot}{\bf A}$:
	\begin{multline}
	\label{Bz(0)}
	B_z(d)-B_z(0)=A_{0y}\int_0^{d} \frac{dx}{\lambda^2(x)}+4\pi M_0\int_0^{d_1} \frac{xdx}{\lambda^2(x)}+\\+4\pi M_0\int_{d_1}^{d} \frac{[d_1 +(x-d_1)\cos \theta]dx}{\lambda^2(x)},
	\end{multline}
	\begin{multline}
	\label{By(0)}
	B_y(d)-B_y(0)=-A_{0z}\int_0^{d} \frac{dx}{\lambda^2(x)}+\\+4\pi M_0\sin \theta \int_{d_1}^{d} \frac{(x-d_1)dx}{\lambda^2(x)}.
	\end{multline}
	
	The first terms in the r.h.s of the equations (\ref{Bz(0)}) and (\ref{By(0)}) can be neglected since they are much smaller than ${\bf B}(0)$. Indeed, these terms are of the order of $[A_{0y(0z)}d]/\lambda^2 \sim (d/\lambda)B_{z(y)}(0) \ll B_{z(y)}(0)$. In the absence of the external magnetic field ${\bf B}(d)=0$ and we immediately find that the components of the magnetic field induced in the superconductor have the form
	\begin{equation}
	\label{B}
	{\bf B}=-4\pi M_0 {\bf Q} e^{x/\lambda_0},
	\end{equation}
	where the components of the vector ${\bf Q}$ have the form $Q_z=(Q_1+Q_2 \cos \theta+Q_3)$ and $Q_y=Q_2 \sin \theta$ with
	
	\begin{equation}
	\begin{gathered}
	\label{Q_def}
	Q_1=\int_0^{d_1} \frac{xdx}{\lambda^2(x)}, \qquad Q_2=\int_{d_1}^{d} \frac{(x-d_1)dx}{\lambda^2(x)},\\
	Q_3=\int_{d_1}^{d} \frac{d_1 dx}{\lambda^2(x)}.
	\end{gathered}
	\end{equation}

	The vector ${\bf Q}$ determines the electromagnetic kernel which controls the magnetic field induced in the superconductor due to the electromagnetic proximity effect.
	
	Our next step is to calculate the components $Q_y$ and $Q_z$. The London penetration depth $\lambda(x)$ is defined by the singlet ($f_s$) and triplet (${\bf f}_t$) components of the anomalous quasiclassical Green function $\hat f(x)=f_s+{\bf f}_t \bm{\sigma}$ which is 2$\times$2 matrix in the spin space ($\bm {\sigma}$ is the vector of Pauli matrices):
	\begin{equation}
	\label{lambda}
	\frac{1}{{{\lambda }^{2}}(x)}=\frac{16{{\pi }^{2}}T\sigma(x) }{{{c}^{2}}}\sum\limits_{\omega_n >0}{\left[|{{f}_{s}(x)}{{|}^{2}}-|{{{\bf f}}_{t}(x)}{{|}^{2}}\right]}.
	\end{equation}
	Here $\omega_n=\pi T (2n+1)$ are the Matsubara frequencies and $\sigma(x)$ is the normal state conductivity. Throughout the paper we put $\hbar=1$.
	
	The anomalous Green functions in the ferromagnets can be obtained from the Usadel equations \cite{Eschrig_PRB}
	\begin{equation}
	\label{Usadel_Eq}
	D\partial_x^2 f_s =2\omega_n f_s +2i {\bf h}{\bf f}_t, \qquad D\partial_x^2 {\bf f}_t =2\omega_n {\bf f}_t +2i {\bf h}f_s,
	\end{equation}
	where ${\bf h}$ is the exchange field and $D$ is diffusion constant.
	
	For simplicity we assume  $\sigma_s \gg \sigma_{f1}$, where $\sigma_s$ and $\sigma_{f1}$ are the conductivities of the S and F$_1$ layers, respectively. This allows imposing the rigid boundary condition for the anomalous Green function at the S/F$_1$ interface: $f_s=f_{s0}=\Delta/ \sqrt{\Delta^2+\omega_n^2}$, ${\bf f}_t=0$, where $\Delta$ is the superconducting gap which is assumed to be real. At the same time, at F$_1$/F$_2$ interface the anomalous Green function should be continuous. The system (\ref{Usadel_Eq}) with the described boundary condition allows the analytical solution which after substitution to Eqs.~(\ref{lambda}) and (\ref{Q_def}) gives the components of the electromagnetic kernel ${\bf Q}$. The resulting expressions for arbitrary angle $\theta$ between magnetic moments are rather cumbersome and, therefore, presented in Appendix I. Keeping in mind the experimental situation relevant to Ref.~\onlinecite{Lee_NatPhys} here we focus on the difference between the cases $\theta=0$ and $\theta=\pi/2$. For simplicity we assume the equal conductivities, diffusion coefficients and the magnitudes of the exchange field in two ferromagnets.
	
	For the parallel orientation of the magnetic moments ($\theta=0$) the kernel ${\bf Q}$ has only one component $Q_z^{\parallel}$ directed along the magnetic moments which coincides with the one for the S/F bilayer with the ferromagnet of the width $d=d_1+d_2$ (see Ref.~\onlinecite{Mironov_APL_2018}):
	\begin{equation}\label{Q_par_res}
	Q_z^{\parallel}=\alpha \sum_{\omega_n>0} f_{s0}^2  {\rm Re} \left[\frac{q^2 d^2+\sinh^2(qd)}{q^2 \cosh^2(qd)}\right],
	\end{equation}
	where  $\alpha=4\pi^2T\sigma/c^2$ and $q=\sqrt{2(\omega_n+ih)/D}$.
	
	\begin{figure}[t!]
		\includegraphics[width=0.8\linewidth]{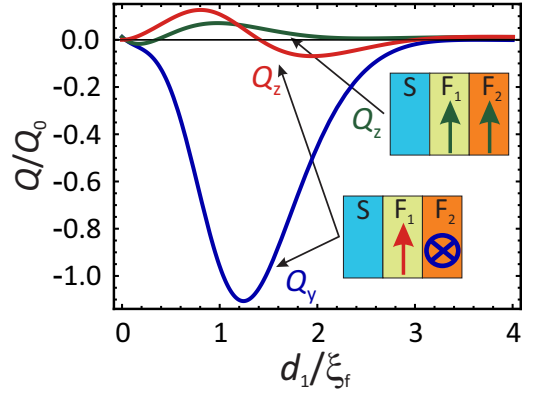}
		\caption{The dependencies of magnetic kernels $Q$ for dirty S/F$_1$/F$_2$ structure on the F$_1$ ferromagnet thickness $d_1$. The green curve corresponds to $Q_z$ for the parallel orientation of magnetic moments, while the red and the blue ones are $Q_z$ and $Q_y$ for perpendicular orientation, respectively. Here $Q_0=16\pi^2T\sigma_f \xi_f^2/c^2$ and we take $\Delta=2\pi T$, $h=50\pi T$, $d_2=10\xi_f$.} \label{Fig:Q_dirty}
	\end{figure}

	For the perpendicular orientation of the magnetic moments ($\theta=\pi/2$) the kernel ${\bf Q}$ and the corresponding spontaneous magnetic field arising in the S layer have both $z$- and $y$-components. To make the expressions for ${\bf Q}$ more transparent we additionally assume $d_1 \sim \xi_f$, $d_2 \gtrsim \xi_n$, where $\xi_f=\sqrt{D/h}$ is the superconducting coherence length in the ferromagnet and $\xi_n=\sqrt{D/T}$ is the normal metal coherence length. In this limit we find: 
	\begin{equation}
	\label{Qy_Perp}
	Q_y^{\perp}=-\alpha \sum_{\omega_n>0} \frac{f_{s0}^2\gamma^2[p^2d_2^2+\sinh^2(pd_2)]}{R^2 p^2 \cosh^2(pd_2)},
	\end{equation}

	\begin{multline}
	\label{Qz_Perp}
	Q_z^{\perp}=\alpha\sum_{\omega_n>0} f_{s0}^2 {\rm Re}\biggl\{ \frac{1}{q^2 \cosh^2 \chi} \biggl[ q^2d_1^2+\\+qd_1 \sinh(2qd_1+2\chi)-\sinh(qd_1+2\chi)\sinh(qd_1)\biggr] \biggr\}.
	\end{multline}
	In these expressions we have introduced the following values: $p=\sqrt{2\omega_n/D}$,
	
	\begin{equation*}
	\gamma={\rm Im}\left\{Q(p,q)\left[\frac{Q(q^*,q^*)}{\cosh(qd_1)}-\frac{Q(q,q^*)}{\cosh(q^*d_1)}\right] \right\},
	\end{equation*}
	
	\begin{equation*}
	R={\rm Re} \left\{Q(p,q^*)\left[Q(q,q)Q(q^*,p)+Q(q^*,q)Q(q,p)\right] \right\},
	\end{equation*}

	\begin{equation*}
	\tanh \chi=\frac{\beta+i\gamma}{R \sinh(qd_1)}-\frac{1}{\tanh(qd_1)},
	\end{equation*}
	where $\beta=2{\rm Re}[Q(p,q^*)]{\rm Re} \left[Q(q,p) / \cosh(q^*d_1) \right]$ and the function $Q(\nu_1,\nu_2)$ depending on two wave-vectors is determined as $Q(\nu_1,\nu_2)=1+[\nu_2 \tanh(\nu_1d_1) \tanh(\nu_2d_2)]/ \nu_1$.
	
	
	\begin{figure}[t!]
		\includegraphics[width=0.8\linewidth]{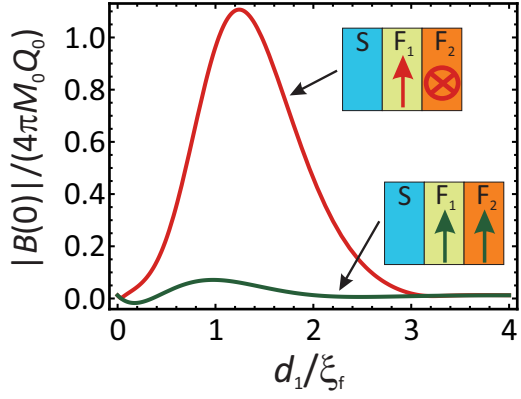}
		\caption{Spontaneous magnetic field at the S/F$_1$ interface induced due to the electromagnetic proximity effect as a function of the F$_1$ layer thickness. Green (red) curve corresponds to the parallel (perpendicular) orientation of the magnetic moments in the ferromagnetic layers.} \label{Fig:B_dirty}
	\end{figure}
	
	The analysis of Eqs.~(\ref{Qy_Perp})-(\ref{Qz_Perp}) shows that the magnetic field induced in the S layer is substantially stronger for the perpendicular orientation of magnetic moments in the F layers compared to the parallel one. Such counterintuitive effect originates from the formation of the long-range spin-triplet superconducting correlations when the two magnetic moments are non-collinear. To illustrate this in more detail let us assume that $d_2\gg \xi_n$ and $h\gg T_c$. For the parallel orientation ($\theta=0$) the superconducting correlations characterized by the component $f_{tz}$ of the anomalous Green function penetrate the ferromagnet over the distance $\sim\xi_f$ and the estimate for the only component of the magnetic kernel ${\bf Q}$ [see Eqs.~(\ref{Q_def}) and (\ref{Q_par_res})] gives 
	$Q_z^{\parallel} \sim (\xi_f /\lambda)^2$. At the same time, for the perpendicular orientation of magnetic moments ($\theta=\pi/2$) the $f_{tz}$ component of the Green function generated in the F$_1$ layer is insensitive to the exchange field in the F$_2$ ferromagnet and the decaying scale of $f_{tz}$ becomes of the order of the normal metal coherence length $\xi_n$ instead of $\xi_f\ll\xi_n$. As a result, the screening parameter $\lambda^{-2}$ stays substantially large in the region of the width $\sim\xi_n$ in the F$_2$ layer which gives $Q_y\sim (\xi_n /\lambda)^2$. This $Q_y$ component of the kernel strongly exceeds the one $Q_z\sim (\xi_f /\lambda)^2$ which is determined only by the value of $Q_1$ for $\theta=\pi/2$.  Note that the same estimates for the kernel components can be also directly extracted from the resulting expressions (\ref{Qy_Perp}) and (\ref{Qz_Perp}). Indeed, for the chosen thicknesses of the F layers the functions $Q(\nu_1,\nu_2)$ and all the values $\gamma$, $\beta$, $R$ are of the order of $1$. Then one sees that each term in the sum (\ref{Qy_Perp}) $\propto p^{-2}\propto \xi_n^2$ while in Eq.~(\ref{Qz_Perp}) it is proportional to $q^{-2}\propto \xi_f^2$ which confirms the above estimate $Q_y/Q_z\sim (\xi_n/\xi_f)^2\gg 1$.

	\begin{figure}[t!]
		\includegraphics[width=0.8\linewidth]{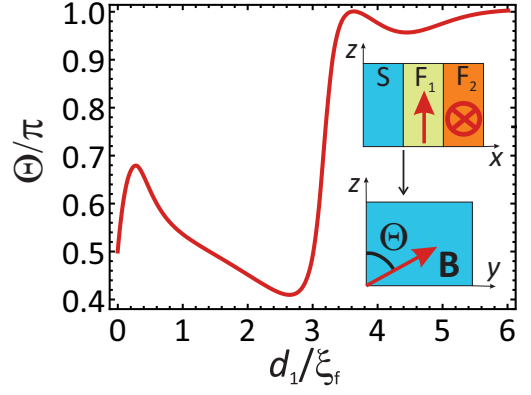}
		\caption{Angle $\Theta$ between this induced magnetic field in the superconductor and the $z$-axis as a function of the thickness $d_1$ of the F$_1$ layer in the case of perpendicular orientation of the magnetic moment in the ferromagnets ($\theta=\pi/2$).} \label{Fig:Angle}
	\end{figure}

	Fig.~\ref{Fig:Q_dirty} shows the dependencies of the different components of the kernel ${\bf Q}$ on the F$_1$ layer thickness for parallel and perpendicular orientations of the magnetic moments in the ferromagnets. One sees that the component $Q_y^{\perp}$ is much more pronounced than both $Q_z^{\parallel}$ and $Q_z^{\perp}$ for $d_1 \sim \xi_f$. As a result the total magnetic field induced inside the superconductors appears to be much larger for $\theta=\pi/2$ compared to the case of collinear magnetic moments $\theta=0$ (see Fig.~\ref{Fig:B_dirty}). Thus, our theory explains the resent puzzling experiments on the Au/Nb/ferromagnet structures where the rotation of the exchange field in the F$_2$ layer from $\theta=0$ to $\theta=\pi/2$ has resulted in the substantial increase of the induced magnetic moment in the superconductor.
	
	Interestingly, at the fixed angle $\theta$ between magnetic moments in ferromagnets the direction of the spontaneous magnetic field inside the superconductor strongly depends on the F$_1$ layer thickness.  In Fig.~\ref{Fig:Angle} we plot the angle $\Theta$ between this induced magnetic field and the $z$-axis as a function of $d_1$ for $\theta=\pi/2$. The complicated oscillatory behavior of $\Theta(d_1)$ demonstrates the rich interference physics associated with the electromagnetic proximity effect in S/F systems.

	\subsection{Clean limit}
	
	The dependence of the spontaneous magnetic field induced in the superconductor on the mutual orientation of the magnetizations in the ferromagnets is more pronounced for clean structures where the Cooper pairs penetrate to the ferromagnet over larger distances and, thus, the proximity effect is stronger. To analyze the electromagnetic proximity effect in this case we consider clean S/F$_1$/F$_2$ structure and calculate the magnetic field induced in the superconductor for the parallel and perpendicular orientations of the magnetic moments in the F$_1$ and F$_2$ layers. Our starting point is again  Eq. (\ref{MaxEq}). However, for clean structures the relation between the Meissner current ${\bf j}_s$ and the vector potential ${\bf A}$ becomes non-local:
	
	\begin{equation}
	\label{js_clean1}
	{\bf j}_s(x)=-\frac{c}{4\pi} \int_{-\infty}^{d} {\bf A}(x')K(x,x')dx'.
	\end{equation}
	Deep inside the superconductor, i.e. for $|x| \gg \xi_0=v_F/T$, the magnetic kernel $K(x,x')$ is local and has the standard London form $K(x,x')=\lambda_0^{-2}\delta(x-x')$ where $\lambda_0$ is the bulk magnetic penetration depth. At the same time, the proximity effect results in the appearance of the London screening in the F layer and the renormalization of the kernel at $x<0$ in the surface layer of the thickness $\sim\xi_0$ near the S/F interface. Since we assume both $d$ and $\xi_0$ to be much less than $\lambda$ we can write down the supercurrent in the following form
	\begin{equation}
	\label{js_clean2}
	{\bf j}_s(x)={\bf j}_{M} -\frac{c}{4\pi} \delta(x) \int_{-\infty}^{d} {\bf A}(x')R(x')dx'.
	\end{equation}
	Here ${\bf j}_{M}=-(c/4\pi)\lambda_0^{-2}{\bf A}\theta(-x)$ is the Meissner current flowing deep inside the superconductor and the correction
	\begin{equation}\label{R_def}
	R(x')=\int_{-\infty}^{d} [K(x,x')-\lambda_0^{-2}\theta(-x) \delta(x-x')]dx,
	\end{equation} 
	is substantially nonzero only in the layer of the thickness $\sim\xi_0$ near the S/F interface [here $\theta(x)$ is Heaviside step function].
	
	The vector-potential in the whole structure has the following form:  ${\bf A}=[A_{0y}(x)+A_{My}]{\bf e_y}+[A_{0z}(x)+A_{Mz}]{\bf e_z}$, where we separate the parts  induced by the magnetizations $A_{My}$ and $A_{Mz}$: $A_{My}=4\pi M_0 x$, $A_{Mz}=0$ for $0<x<d_1$ and $A_{My}=4\pi M_0 d_1 +4\pi M_0(x-d_1)\cos \theta$, $A_{Mz}=-4\pi M_0(x-d_1)\sin \theta$ for $d_1<x<d_2$. To calculate the supercurrent we substitute these expressions for the vector-potential into Eq.(\ref{js_clean2}). Inside the superconductor the vector potential components $A_{0y}(x)$ and $A_{0z}(x)$ decay at the distances $\sim\lambda\gg\xi_0$ from the S/F interface and, thus, they can be treated as  constant in the surface region of the non-locality. Then the supercurrent reads
	
	\begin{equation}
	\label{js_fen}
	{\bf j}_s={\bf j}_{M}-\frac{c}{4\pi}[(P_z+4\pi M_0 Q_z){\bf e_y}+(P_y-4\pi M_0 Q_y){\bf e_z}]\delta (x),
	\end{equation}
	where we have introduced the following values:
	\begin{equation}
	P_z=A_{0y} \int_{-\infty}^{d}R(x')dx', ~~~ P_y=A_{0z} \int_{-\infty}^{d}R(x')dx',
	\end{equation}
	\begin{equation}
	\label{Qz_clean_fen}
	Q_z=\int_0^{d_1}x'R(x')dx' + \int_{d_1}^{d}[d_1+(x'-d_1)\cos \theta]R(x')dx',
	\end{equation}
	\begin{equation}
	\label{Qy_clean_fen}
	Q_y=\int_{d_1}^{d}(x'-d_1)\sin \theta R(x')dx'.
	\end{equation}
	In the above expressions the terms proportional to $P_z$ and $P_y$ give rise to the small renormalization of the screening current near the S/F$_1$ interface and can be neglected due to the assumptions $d\ll\lambda$ and $\xi_0\ll\lambda$.
	
	Solving Eq.(\ref{MaxEq}) with the supercurrent ${\bf j}_s$ satisfying Eq.(\ref{js_fen}) we obtain that the induced magnetic field in the S layer has the form (\ref{B}) where $Q_z$ and $Q_y$ are defined by Eqs.~(\ref{Qz_clean_fen}) and (\ref{Qy_clean_fen}), respectively. To perform the quantitative analysis of the induced magnetic field we now find the values $Q_z$ and $Q_y$ microscopically. To do this we calculate the current ${\bf j}_s$ within the Eilenberger formalism and comparing it with the phenomenological expression (\ref{js_fen}) obtain the desired expressions for the magnetic kernel componets. 
	
	The superconducting current ${\bf j}_s$ flowing along the S/F$_1$ interface is determined by the singlet component of the normal Green function $g_s$ as
	\begin{equation}
	\label{j}
	{\bf j}_s=-4\pi e \nu_0 T\sum\limits_{\omega >0} \langle {\bf v}_F{\rm Im}[g_s] \rangle.
	\end{equation}
	Here ${\bf v}_F=v_x {\bf e}_x+v_y {\bf e}_y+v_z {\bf e}_z$ is the vector of quasiparticle velocity along the quasiclassical trajectory which has the components $v_x=v_F \sin \theta \cos \varphi$, $v_y=v_F \sin \theta \sin \varphi$, and $v_z=v_F \cos \theta$; $\nu_0$ is the density of states at the Fermi level per unit spin projection; and the brackets denote the averaging over the Fermi surface: $\langle ...\rangle =(4\pi)^{-1}\int\limits_{0}^{\pi }\sin \theta d\theta \int\limits_{0}^{2\pi }{...\,\,d\varphi }$.
	
	To calculate the function $g_s$ we solve the Eilenberger equations along the quasiclassical trajectories \cite{Eschrig_PRB}. Inside the superconducting layer they take the form
	\begin{equation}
	\label{Eil_S_g}
	{\bf v}_F\hat {\bm \nabla}\hat g=\Delta^* \hat f -\Delta \hat f^{\dagger}
	\end{equation}
	\begin{equation}
	\label{Eil_S_f}
	{\bf v}_F\hat{\bf D}\hat f=-2\omega_n \hat f+2\Delta \hat g, \qquad {\bf v}_F\hat{\bf D}^*\hat f^{\dagger}=2\omega_n \hat f^{\dagger}-2\Delta^* \hat g,
	\end{equation}
	where $\hat g=g_{s}+{\bf g}_{t}\bm{\sigma}$ is the normal Green function and $\hat f=f_{s}+{\bf f}_{t}\bm{\sigma}$, $\hat f^{\dagger}=\tilde{f}_{s}+\tilde{{\bf f}}_{t}\bm{\sigma}$ are the anomalous ones, and $\hat {\bf D}={\bm \nabla}+2ie{\bf A}/c$ is the gauge-invariant momentum operator ($e>0$).
	
	In what follows we focus on the solution of the above equations in the surface region of the thickness $\sim\xi_0$ near the S/F interface, where the surface current gives rise to the significant renormalization of the electromagnetic response. Since $\xi_0 \ll \lambda$ one can neglect the spatial variation of the vector potential components $A_{0y}$ and $A_{0z}$ in the region under consideration and assume them to be constant. In this case the effect of $A_{0y}$ and $A_{0z}$ on the Green functions can be taken into account through the renormalization of Matsubara frequencies $\tilde {\omega}_n=\omega_n + i eA_{0y}v_y/c+ieA_{0z}v_z/c$.

	\begin{figure}[t!]
		\includegraphics[width=0.8\linewidth]{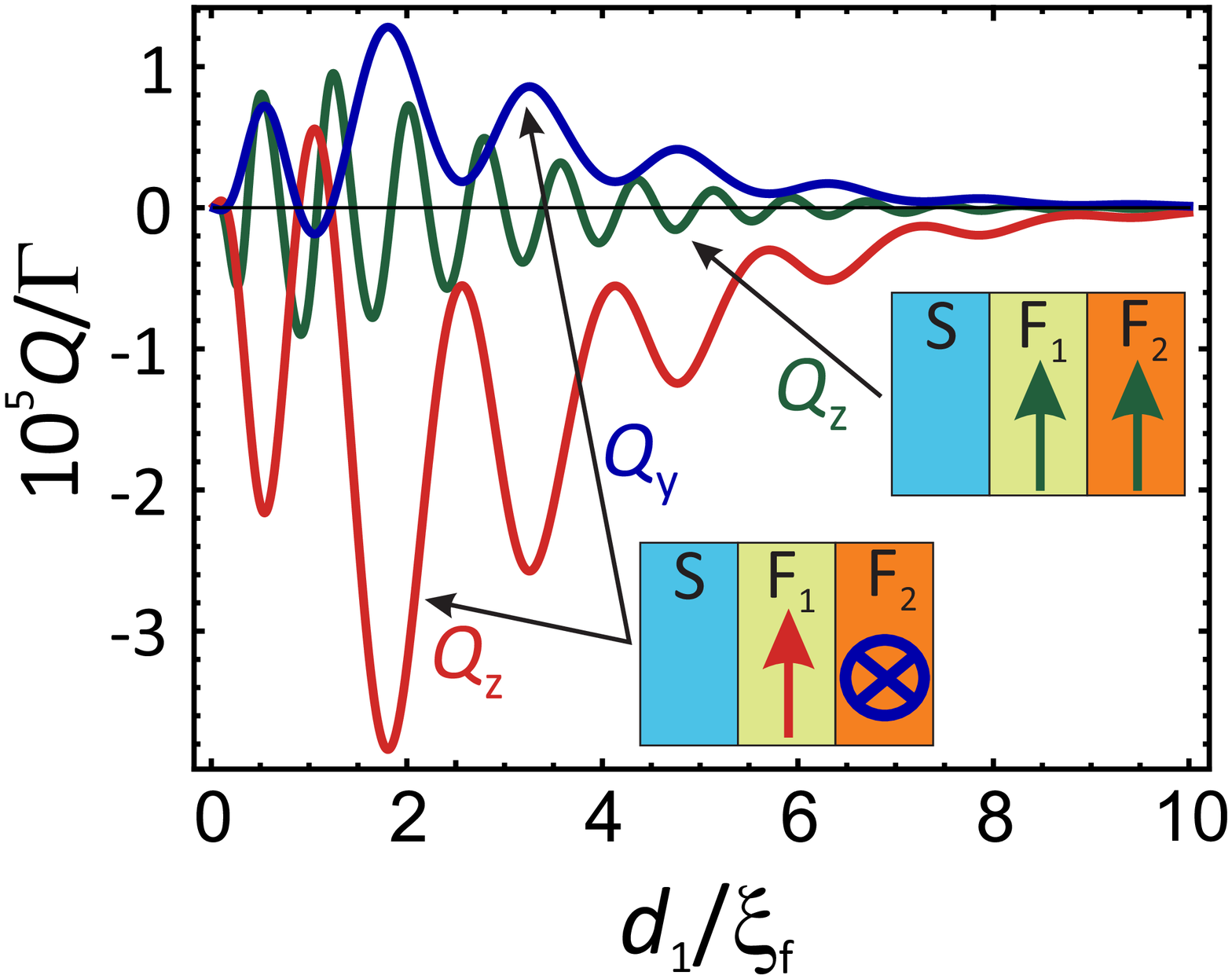}
		\caption{The dependencies of magnetic kernels $Q$ for clean S/F$_1$/F$_2$ structure with identical thicknesses of the ferromagnets, i.e. $d_1=d_2$, on $d_1$. Here $\xi_f=v_F/h$. The green curve corresponds to $Q_z$ for parallel orientation of magnetic moments, while the red and blue ones are $Q_z$ and $Q_y$ for perpendicular orientation, respectively. The parameters are $\Delta=2\pi T$, $h=10\pi T$. {\bf Here $\Gamma=8\pi^2 e^2 \nu_0 v_F^4/(c^2 T^2)$.}} \label{Fig:Q_clean}
	\end{figure}

	Inside the ferromagnets the Eilenberger equations for the anomalous Green functions read \cite{Eschrig_PRB}
	\begin{equation}
	\begin{gathered}
	\label{Eil_F_f}
	{\bf v}_F\hat{\bf D}\hat f=-2\omega_n \hat f-i{\bf h}\bm{\sigma}f-i\hat f{\bf h}\bm{\sigma}, \\
	{\bf v}_F\hat{\bf D}^*\hat f^{\dagger}=2\omega_n \hat f^{\dagger}+i{\bf h}\bm{\sigma}\hat f^{\dagger}+i\hat f^{\dagger}{\bf h}\bm{\sigma}.
	\end{gathered}
	\end{equation}
	
	Our further strategy is to find the solutions of the Eilenberger equations in the S, F$_1$ and F$_2$ layers separately and then match them in the interfaces with the boundary conditions. Taking the second derivative of Eq. (\ref{Eil_S_g}) and using Eq. (\ref{Eil_S_f}) we obtain the equation which contains only the normal Green function inside the superconductor (see Appendix~ \ref{App_Clean} for details).  The solution of this equation is straightforward. Next we substitute the obtained function $\hat g$ into Eq. (\ref{Eil_S_f}) and find the anomalous Green functions $\hat f$ and $\hat f^{\dagger}$. As a next step we find the anomalous Green functions in the ferromagnets. For this purpose we first eliminate the vector-potential from the equations  Eq. (\ref{Eil_F_f}) by means of the gradient transformation and then solve the obtained equations which have the standard form. Finally, we match the anomalous Green functions in the S/F$_1$ and F$_1$/F$_2$ interfaces and find all unknown constants. For details of the calculations see Appendix~\ref{App_Clean}. The final expressions for the singlet components of the normal Green function in the superconductor for parallel ($g_{s}^{\parallel}$) and perpendicular ($g_{s}^{\perp}$) orientations of the magnetic moments in the F layers have the following form:
	\begin{multline}
	\label{gs}
	g_{s}^{\parallel(\perp)}(x)=\frac{\tilde{\omega}_n}{\tilde{\Omega}_n}+\frac{2\Delta^2}{\tilde{\Omega}_n |v_x|}e^{k_nx}\times \\ \times \frac{(k_n+\kappa_n {\rm tanh}\varphi_n ){\rm tanh}\varphi_n +(\kappa_n +k_n{\rm tanh}\varphi_n )t_{\parallel(\perp)}}{(k_n+\kappa_n {\rm tanh}\varphi_n )^2+{(\kappa_n +k_n{\rm tanh}\varphi_n )}^2 t_{\parallel(\perp)}},
	\end{multline}
	where $\tilde{\Omega}_n=\sqrt{\tilde{\omega}_n^2+\Delta^2}$, $k_n=2\tilde{\Omega}_n/|v_x|$, $\kappa_n=2\tilde{\omega}_n/|v_x|$, $t_{\parallel}={\rm tan}^2(\psi_1+\psi_2)$, $t_{\perp}=({\rm tan}^2\psi_1+{\rm tan}^2\psi_2)$,  $\psi_j=2hd_j/v_x$, and
	\begin{equation}
	\varphi_n =\kappa_n d+\frac{2ie{v_y}}{c |v_x|}\int\limits_0^d A_{My}(x')dx'+\frac{2ie{v_z}}{c |v_x|}\int\limits_0^d A_{Mz}(x')dx'.
	\end{equation}
	At the same time, inside the ferromagnets both $g_{s}^{\parallel}$ and $g_{s}^{\perp}$ are uniform and equal to  $g_{s}^{\parallel}(0)$ and $g_{s}^{\perp}(0)$, respectively.
	
	For the parallel orientation of the magnetic moments in the two F layers the function $g_{s}^{\parallel}$ for the S/F$_1$/F$_2$ structure is the same as the singlet component of the normal Green function in the S/F bilayer where the thickness of the ferromagnet equals to $d$. Thus, for $\theta=0$ we have $Q_y^{\parallel}=0$ while $Q_z^{\parallel}$ coincides with one obtained in Ref.[\onlinecite{Mironov_APL_2018}].
	
	For the perpendicular orientation of the magnetic moments ($\theta=\pi/2$) to calculate the supercurrent we expand the function $g_s$ up to the first order over the vector-potential and then substitute it into Eq.(\ref{j}). Further, it is convenient to represent the supercurrent as ${\bf j}_s={\bf j}_M+{\bf j}^{surf}$, where  ${\bf j}_M$ is the standard Meissner current flowing in the bulk superconductor and ${\bf j}^{surf}$ is the surface correction arising due to the proximity effect. Neglecting the small renormalization of the ${\bf j}^{surf}$ caused by ${\bf A}_0$ we can write it down in the following form
	
	\begin{equation}
	\label{jysurf}
	j_y^{surf}(x)=-\frac{2\Gamma (d_1^2+2d_1 d_2)cTM_0}{\xi_0^2 v_F^2}\sum\limits_{n\ge 0}\frac{\Delta^2}{\Omega_n^2}\left\langle \left. \frac{v_y^2}{|v_x|}P_n e^{k_{0n} x}\right\rangle  \right.,
	\end{equation}
	
	\begin{equation}
	\label{jzsurf}
	j_z^{surf}(x)=\frac{2\Gamma (d_2^2-2d_1 d_2)cTM_0}{\xi_0^2 v_F^2}\sum\limits_{n\ge 0}\frac{\Delta^2}{\Omega_n^2}\left\langle \left. \frac{v_z^2}{|v_x|}P_n e^{k_{0n} x}\right\rangle  \right..
	\end{equation}
	Here $\Gamma=8\pi^2 e^2 \nu_0 v_F^4/(c^2 T^2)$, $\Omega_n=\sqrt{\omega_n^2 +\Delta^2}$, $k_{0n}=2\Omega_n /|v_x|$,  and
	
	\begin{multline}
	P_n=\frac{1}{\sinh^2(\kappa_{0n} d)}
	\biggl\{ \frac{[\rho+S(\rho)]\tanh(\kappa_{0n} d)+t_{\perp}}{S^2(\rho)+\rho^2 S^2(\rho^{-1})t_{\perp}}-\\-\frac{2\rho[S(\rho)\tanh(\kappa_{0n} d)+\rho S(\rho^{-1})t_{\perp}][S(\rho)+S(\rho^{-1})t_{\perp}]}{[S^2(\rho)+\rho^2 S^2(\rho^{-1})t_{\perp}]^2} \biggr\},
	\end{multline}
	where $\kappa_{0n}=2\omega_n/|v_x|$, $S(x)=x+\coth(\kappa_{0n} d)$, $\rho=\omega_n /\Omega_n$. In the ferromagnets the surface current is equal to ${\bf j}^{surf}(0)$. Finally, we integrate ${\bf j}^{surf}$ over $x$ and comparing obtained expression with Eq.(\ref{js_fen}) find the  kernels $Q_z$ and $Q_y$ for the perpendicular magnetic configuration:
	\begin{equation}
	\label{Qz_clean}
	Q_z^{\perp}=\frac{\Gamma (d_1^2+2d_1 d_2)}{\xi_0^2}\sum\limits_{n\ge 0}\frac{T\Delta^2}{\Omega_n^3}\left\langle \left. \frac{v_y^2}{v_F^2}\left(1+k_{0n} d \right)P_n\right\rangle  \right.,
	\end{equation}
	
	\begin{equation}
	\label{Qy_clean}
	Q_y^{\perp}=\frac{\Gamma (d_2^2-2d_1 d_2)}{\xi_0^2}\sum\limits_{n\ge 0}\frac{T\Delta^2}{\Omega_n^3}\left\langle \left. \frac{v_z^2}{v_F^2}\left(1+k_{0n} d \right)P_n\right\rangle  \right..
	\end{equation}
	
	For the clean S/F$_1$/F$_2$ structure with perpendicular orientation of magnetic moments and identical thicknesses of the ferromagnetic layers, i.e. $d_1=d_2$ the typical dependencies  $Q_z^{\perp}(d_1)$ and $Q_y^{\perp}(d_1)$ are shown in Fig.~\ref{Fig:Q_clean}. For comparison we also plot $Q_z^{\parallel}(d_1)$ for the same structure with parallel magnetic configuration. As  expected, the induced magnetic field in the superconductor is more pronounced for $\theta=\pi/2$ than for $\theta=0$. Another distinctive feature of these dependencies is the oscillatory behavior of the corresponding magnetic kernels. The dependencies $Q_z^{\perp}(d_1)$ and $Q_y^{\perp}(d_1)$ demonstrate oscillations with the period of the order of $\xi_h =v_F/h$, while the period of the oscillations of $Q_z^{\parallel}(d_1)$ is two times smaller. The change in the $Q_z$ ($Q_y$) sign causes the corresponding changes in the sign of $B_z$ ($B_y$) component of the induced magnetic field in the superconductor. Thus, depending on $d_1$ the $B_z$ and $B_y$ components of the field can be either parallel or antiparallel to the corresponding magnetization components. At $d\ll\xi_0$ the envelopes of all dependencies increases as functions of $d_1$ being proportional to $ (d/\xi_0)^2$ at $d \ll \xi_0$ while at $d \gg \xi_0$ they decay exponentially. In contrast to $Q_z^{\parallel}(d_1)$ the dependencies $Q_z^{\perp}(d_1)$ and $Q_y^{\perp}(d_1)$ are not symmetric with respect to the $x$-axis. Moreover, $Q_{z\perp}$ and $Q_{y\perp}$ have opposite sign. Thus, if $B_y$ is parallel to ${\bf M}_2$, then $B_z$  is antiparallel to ${\bf M}_1$.

	\section{Shift in Fraunhofer critical current oscillations in S1/I/S2/F structure}
	
	\begin{figure}[t!]
		\includegraphics[width=0.8\linewidth]{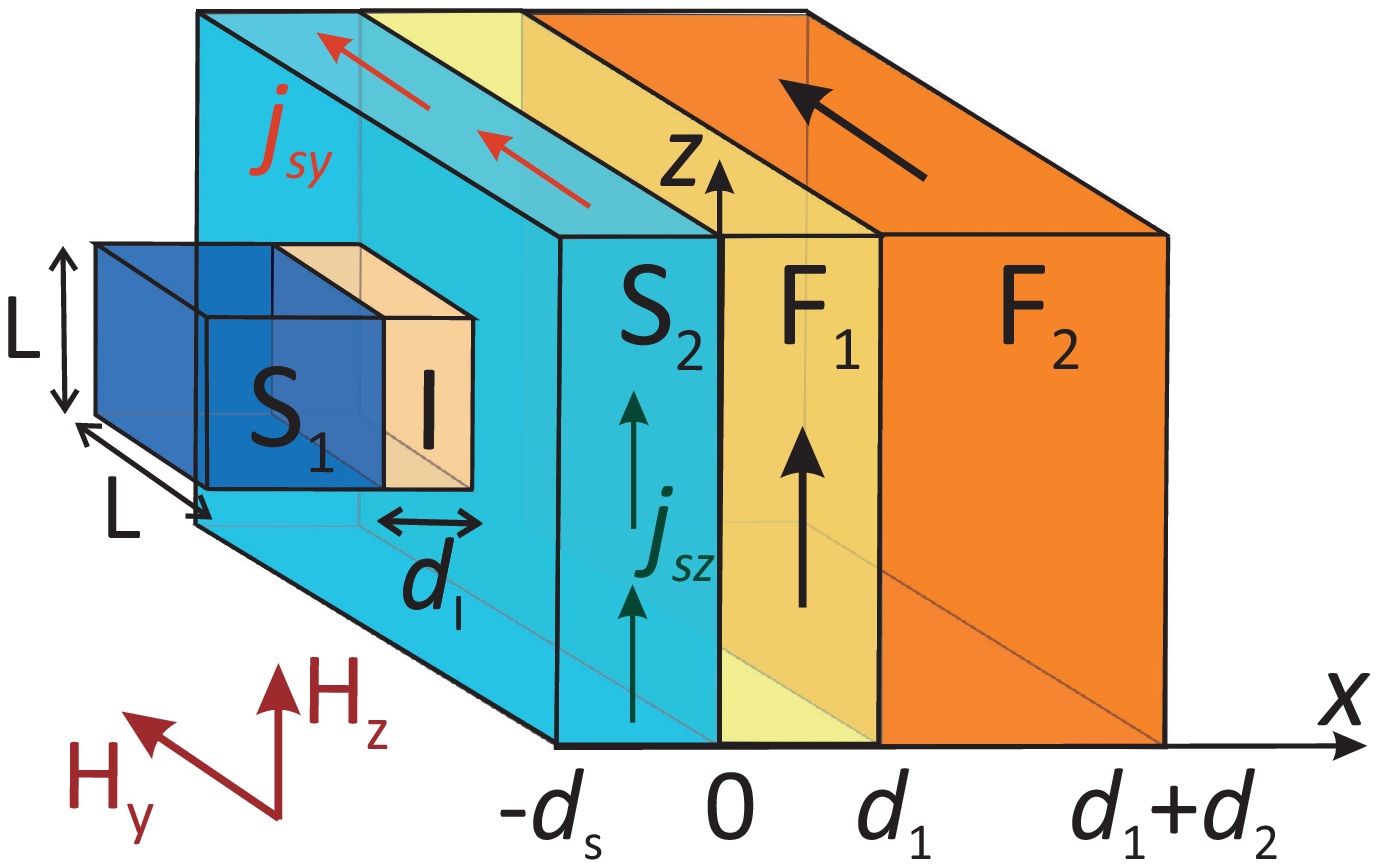}
		\caption{The sketch of the Josephson junction with one superconducting electrode covered by the composite ferromagnet. } \label{Fig:JJ}
	\end{figure}

	The electromagnetic proximity effect also manifest itself in the shift of the Fraunhofer dependence of the critical current vs. the external magnetic field in the Josephson junction with one electrode covered by composite ferromagnetic layer. This provides the possibility to determine experimentally both the modulus and the direction of the magnetic field induced in the superconductor. Indeed, let us consider the junction (see Fig.{\ref{Fig:JJ}}) where S$_2$ electrode with the thickness of the order of $\lambda$ is covered by the composite ferromagnet.  The other superconducting electrode S$_1$ has the thickness $d_{s1}\lesssim\lambda$ and the square cross-section with the side $L$ satisfying the condition $\lambda < L < \lambda_J$, where $\lambda_J$ is Josephson penetration depth. Since $L < \lambda_J$ the probe S$_1$  does not perturb the measured magnetic field distribution. The insulating layer I has the thickness $d_I$. If the junction is placed into the external magnetic field ${\bf H}=H_z {\bf e}_z+H_y {\bf e}_y$, the Josephson phase $\phi$ obeys the following equation \cite{Mironov_AJE}
	
	\begin{equation}
	\label{FPE}
	\nabla \phi =\frac{2\pi}{\Phi_0}\left[\frac{4\pi \lambda_0^2}{c}({\bf j}^+_s -{\bf j}^-_s)+({\bf A}^+_s -{\bf A}^-_s)\right],
	\end{equation}
	where the signs "$+$"("$-$")  correspond to $x=-d_s$ ($x=-d_s-d_I$) and $\Phi_0$ is superconducting flux quantum. Since the thickness of the insulating layer is rather small we can approximately write down $A_y^+ - A_y^- \approx H_z d_I$ and $A_z^+ - A_z^- \approx -H_y d_I$. The supercurrent ${\bf j}_s^-$  at the S$_1$/I interface has the standard form ${\bf j}^{-}_{s}=\left[ \hat{\bf e}_x\times {\bf H}\right]\tanh\left(d_{s1}/2\lambda_0\right)c/(4\pi \lambda_0)$ ($\hat{\bf e}_x$ is the unit vector along the $x$ axis), while ${\bf j}_s^+$ is renormalized due to the presence of the spontaneous magnetic field inside the S$_2$ superconductor. To calculate ${\bf j}_s = [c/(4\pi)]\rm{rot}{\bf B}$ inside the S$_2$ layer we solve the London equation (\ref{MaxEq}) with the boundary conditions ${\bf B}(0)={\bf H}-4\pi M_0 {\bf Q}$ and ${\bf B}(-d_s)={\bf H}$. The former condition follows from Eqs.(\ref{Bz(0)}), (\ref{By(0)}). We obtain for the supercurrent:
	
	\begin{equation}
	j^+_{s\alpha}=s_{\alpha}\frac{c}{4\pi\lambda_0}\left\{H_{\beta}\tanh\left(\frac{d_s}{2\lambda_0}\right)+\frac{4\pi M_0 Q_{\beta}}{\sinh(d_s/\lambda_0)}\right\},
	\end{equation}
	Here if $\alpha=y$ ($\alpha=z$), then $\beta=z$ ($\beta=y$), $s_y=1$, $s_z=-1$. Substituting the found currents into Eq.(\ref{FPE}) we immediately obtain the  Josephson phase $\phi(y,z)=n_z y-n_y z+C$. Here $C$ is the constant and the coefficients are
	\begin{equation}\label{n_def}
	n_{z(y)}=\frac{2\pi}{\Phi_0}\left[H_{z(y)}(d_I+2\tilde{\lambda})+\frac{4\pi M_0 Q_{z(y)}\lambda_0}{\sinh(d_s/\lambda_0)}\right],
	\end{equation}
	where $2\tilde{\lambda}=\lambda_0\{\tanh[d_{s1}/(2\lambda_0)]+\tanh[d_s/(2\lambda_0)]\}$ \cite{Barone}. Note that Eq.~(\ref{n_def}) is valid only for $d_s$ values which are much larger than the thickness of the region near the S/F interface where the magnetic kernel substantially deviates from the London one. Exactly this thickness cuts the divergence of Eq.~(\ref{n_def}) at $d_s\to 0$ (see the detailed discussion in Sec. \ref{Sec_FSF}). 
	
	Finally, we find the total current through the junction integrating the current density $j_x(y,z)=j_c\sin(n_z y-n_y z+C)$:
	
	\begin{equation}
	I= -j_c L^2 \cos C \frac{\sin(n_z L/2)}{(n_z L/2)} \frac{\sin(n_y L/2)}{(n_y L/2)}.
	\end{equation}

	\begin{figure}[t!]
		\includegraphics[width=0.7\linewidth]{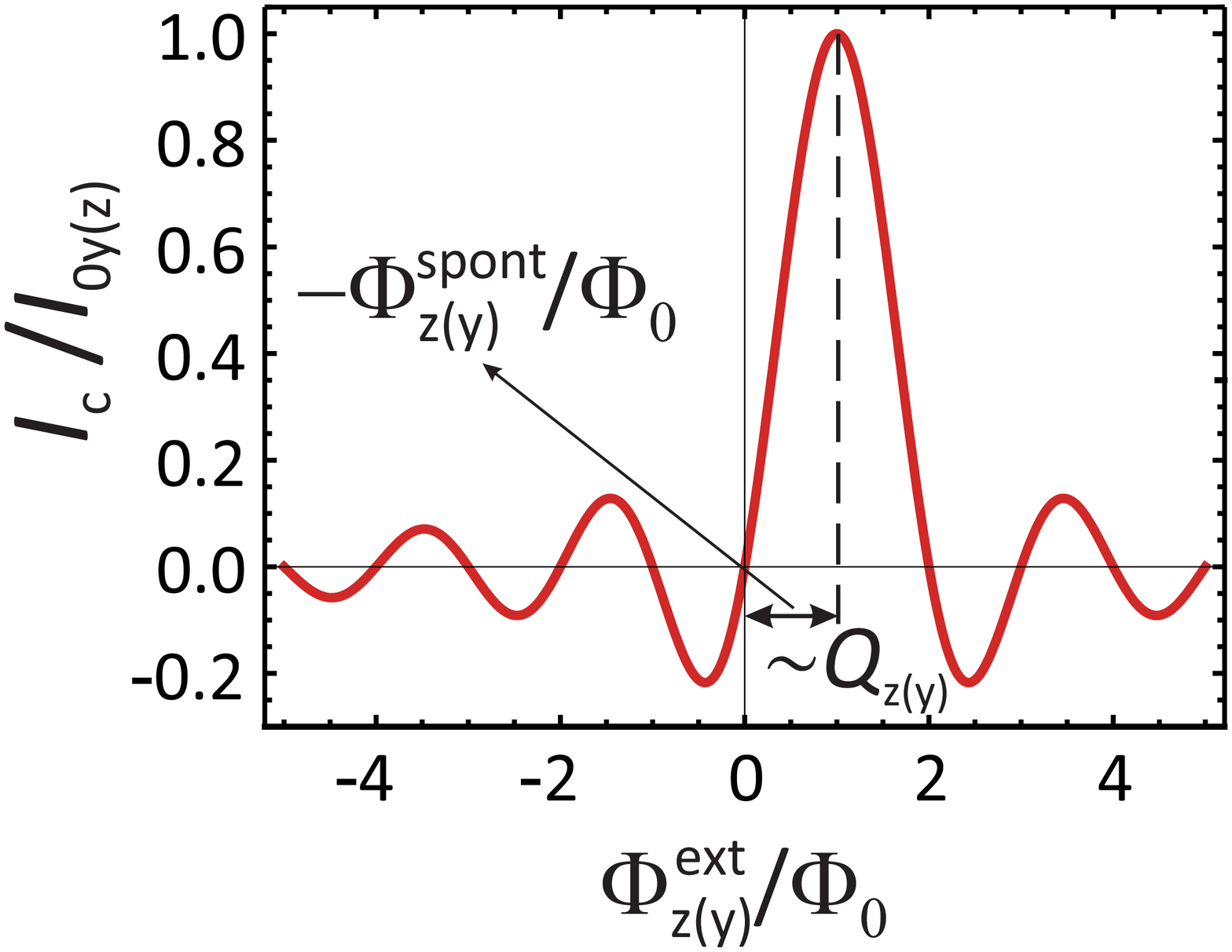}
		\caption{Shift in the Fraunhofer critical current oscillations in the S$_1$/I/S$_2$/F Josephson junction caused by the electromagnetic proximity effect. Here $I_{0y(z)}=[j_c L^2\Phi_0]/[\pi(\Phi_{y(z)}^{ext}+\Phi_{y(z)}^{spont})]\sin\left[\pi(\Phi_{y(z)}^{ext}+\Phi_{y(z)}^{spont}) / \Phi_0\right]$.} \label{Fig:Fraunhofer}
	\end{figure}

	Thus, the maximal current $I_c$ through the junction has the following form
	
	\begin{multline}
	\label{JC}
	I_c=j_c L^2 \frac{\Phi_0^2}{\pi^2(\Phi_y^{ext}+\Phi_y^{spont})(\Phi_z^{ext}+\Phi_z^{spont})}  \times \\ \times \sin\left[ \frac{\pi(\Phi_y^{ext}+\Phi_y^{spont})}{\Phi_0}\right] \sin\left[\frac{\pi(\Phi_z^{ext}+\Phi_z^{spont})}{\Phi_0}\right]
	\end{multline}
	where $\Phi_{y(z)}^{ext}=H_{y(z)}L(d_I+2\tilde{\lambda})$ and $\Phi_{y(z)}^{spont}=[4\pi M_0 Q_{y(z)}\lambda_0 L]/[\sinh(d_s/\lambda_0)]$ are  the magnetic fluxes generated by the external and spontaneous magnetic fields, respectively. If one neglect the later contribution, then the Eq.~(\ref{JC}) takes the standard form for the rectangular junction \cite{Barone}.
	
	One can see, that both dependencies $I_c(H_y)$ and $I_c(H_z)$ are sfifted by the values proportional to the kernels $Q_y$ and $Q_z$, respectively. Thus, sequentially directing the external magnetic field along two perpendicular components of the magnetization of composite ferromagnet and measuring the Fraunhofer dependencies of the critical current one can independently obtain both components of the magnetic field induced in the superconductor due to the electromagnetic proximity effect. Therefore, this technique provides the possibility to determine experimentally both the modulus and direction of the spontaneous magnetic field.

	\section{Long ranged superconductivity control of the magnetic state in F1/S/F2 structures}\label{Sec_FSF}

\begin{figure}[t!]
		\includegraphics[width=1\linewidth]{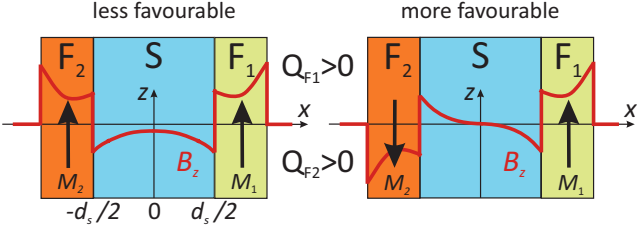}
		\caption{The magnetic states and the corresponding magnetic field profiles in different F$_1$/S/F$_2$ structures. The electromagnetic proximity effect controls the magnetic state of two ferromagnets: in the case when electromagnetic kernels $Q_{F_1}$ and  $Q_{F_2}$ of the F$_1$ and F$_2$ layers are positive antiparallel magnetic configuration is more favorable.} \label{Fig:FSF}
	\end{figure}

	The considered electromagnetic proximity effect mays also play a role in the control of the mutual orientation of the ferromagnetic moments in F/S/F spin valve. Let us consider a simple model where the superconductor S with the thickness $d_s \sim \lambda$ is placed between two ferromagnets F$_1$ and F$_2$ with the thicknesses $d_1 \sim \xi_f \ll \lambda$, $d_2 \sim \xi_f \ll \lambda$ and the magnetizations ${\bf M}_1=M_1 {\bf e}_z$, ${\bf M}_2=M_2 {\bf e}_z$  the  electromagnetic proximity effect causes the orientation of magnetic moments ${\bf M}_1$ and ${\bf M}_2$ (see Fig.{\ref{Fig:FSF}}). If the currents inside both ferromagnets are diamagnetic, then antiparallel magnetic configuration is more favorable. To show it directly we now consider the simple model in which the F$_1$ (F$_2$) ferromagnet is characterized by the London penetration depth $\lambda_1$ ($\lambda_2$) caused by the penetration of Cooper pairs from the superconductor. At first, we find the distribution of the vector potential  ${\bf A}=A {\bf e}_y$ in the whole structure using Eq.(\ref{MaxEq}) and Eq.(\ref{LEq}). As before, inside the superconductor we assume $\lambda (x)=\lambda_0={\rm const}$ and obtain $A_s(x)=K_1\cosh(x/\lambda_0) +K_2 \sinh(x/\lambda_0)$. The vector-potential inside F$_1$ (F$_2$) layer takes the form
	
	\begin{multline}
	A_{F_{1(2)}}=N_{1(2)}\cosh\left(\frac{x\mp L_{1(2)}}{\lambda_{1(2)}}\right)+\\+\frac{4\pi M_{1(2)}\lambda_{1(2)}}{\cosh[L_{1(2)}/\lambda_{1(2)}]} \sinh\left(\frac{x}{\lambda_{1(2)}}\right),
	\end{multline}
	where $L_{1(2)}=d_{1(2)}+d_s/2$. Using the continuity of ${\bf A}$ and $({\bf B}-4\pi {\bf M})$ at the interfaces $x=\pm d_s/2$ we find the coefficients:

	\begin{multline*}
	\begin{gathered}
	K_1=\frac{4\pi(\tilde{M}_1 \lambda_1 r_2 - \tilde{M}_2 \lambda_2 r_1)}{(r_1 v_2 + r_2 v_1)},\\
	K_2=\frac{4\pi(\tilde{M}_1 \lambda_1 v_2 + \tilde{M}_2 \lambda_2 v_1)}{(r_1 v_2 + r_2 v_1)},\\
	N_{1(2)}=\frac{1}{\cosh(d_{1(2)}/\lambda_{1(2)})}\biggl[K_1 \cosh\left(\frac{d_s}{2\lambda_0}\right) \pm \\ \pm K_2 \sinh\left(\frac{d_s}{2\lambda_0}\right)  \mp \frac{4\pi M_{1(2)}\lambda_{1(2)}}{\cosh[L_{1(2)}/\lambda_{1(2)}]} \sinh\left(\frac{d_s}{2\lambda_{1(2)}}\right)\biggr]
	\end{gathered}
	\end{multline*}
	where $\tilde{M}_{1(2)}=M_{1(2)}[1-\cosh(d_{1(2)}/\lambda_{1(2)})]$, $r_{1(2)}=\sinh(d_s/2\lambda_0)\sinh(d_{1(2)}/\lambda_{1(2)})+[\lambda_{1(2)} /\lambda_0]\cosh(d_s/2\lambda_0)\cosh(d_{1(2)}/\lambda_{1(2)})$ and $v_{1(2)}=\cosh(d_s/2\lambda_0)\sinh(d_{1(2)}/\lambda_{1(2)})+[\lambda_{1(2)} /\lambda_0]\sinh(d_s/2\lambda_0)\cosh(d_{1(2)}/\lambda_{1(2)})$.

	Now we calculate the free energy per unit area
	$F=\int \left[({\bf B}-4\pi {\bf M})^2/(8\pi) + {\bf A}^2/(8\pi \lambda^2)\right]dx$ which can be rewritten as follows:
	
	\begin{multline*}
	F=-\frac{1}{2}M_1 \left[A_{F_1}(L_1)- A_{F_1}(d_s/2)\right]+\\+\frac{1}{2}M_2 \left[A_{F_2}(-L_2)- A_{F_2}(-d_s/2)\right]+2\pi^2 (M_1^2 d_1+M_2^2 d_2).
	\end{multline*}
	
	Substituting the found vector-potentials we obtain:
	\begin{equation}
	\label{FSF_FE_exact}
	F=F_0 + \frac{16\pi \lambda_1 \lambda_2  M_1 M_2}{\lambda_0 (v_1 r_2+v_2 r_1)}\sinh^2\left(\frac{d_1}{2\lambda_1}\right) \sinh^2\left(\frac{d_2}{2\lambda_2}\right),
	\end{equation}
	where $F_0$ is the part which does not depend on whether the magnetic moments in the ferromagnets are parallel or antiparallel to each other. Since  $v_{1,2}>0$ and  $r_{1,2}>0$  the antiparallel magnetic configuration is more favorable (see Fig.{\ref{Fig:FSF}}). In the case $\lambda_1=\lambda_2=\lambda_0$ the expression (\ref{FSF_FE_exact}) takes the following transparent form:
	
	\begin{equation}
	\label{FSF_FE}
	F=F_0 + \frac{\pi \lambda_0  M_1 M_2}{\sinh[(d_s+d_1+d_2)/\lambda_0]}\left(\frac{d_1}{\lambda_1}\right)^2\left(\frac{d_2}{\lambda_2}\right)^2.
	\end{equation}

	In general case with a relatively thick ferromagnetic layers we may expect that qualitatively the behavior of the free energy will be similar to (\ref{FSF_FE}), but with the parameters $(d_i/\lambda_i)^2$ replaced in the dirty limit by the corresponding kernels $Q_{F_i} \sim (\xi_{F_i}/\lambda_i)^2$, where $i=1,2$.  Interestingly, the exchange interaction also contribute to the antiferromagnetic ordering of the magnetizations in such structures \cite{deGennes}. Recently it has been demonstrated the possibility to switch the F/S/F spin valve into antiferromagnetic state by superconductivity \cite{Zhu}. This exchange mechanism of switching may operate for S layer thicknesses of the order of superconducting coherence length $\xi_0$. For such thicknesses the discussed electromagnetic mechanism gives greater contribution than the exchange one if the magnetizations satisfy the condition $M_i> (\lambda_i/\xi_f)^2 H_{c1}$, where $H_{c1}$ is the lower critical field. On the other hand the electromagnetic mechanism is a long-ranged and should dominate for S layer thickness larger than $\xi_0$. Note that in the case of diamagnetic currents inside F layers the electromagnetic mechanism favors the antiferromagnetic ordering while in general case one can expect that the ground state configuration is determined by the signs of the electromagnetic kernels which depends on their thicknesses (see Fig.\ref{Fig:Q_dirty}).

	\section{Conclusion}
	
	To sum up, we have developed the theory of the electromagnetic proximity effect in S/F$_1$/F$_2$ structure both in dirty and clean limits. We have demonstrated that in dirty case the magnetic field ${\bf B}$ induced in the superconductor is significantly larger for the perpendicular magnetic configuration ($\perp$) (the magnetization ${\bf M}_1$ in the F$_1$ ferromagnet coincides with $z$-axis, while the magnetization ${\bf M}_2$ in the F$_2$ layer is directed along $y$-axis ) in comparison with the parallel one ($\parallel$) (both ${\bf M}_1$ and ${\bf M}_2$ coincide with $z$-axis), see Fig.\ref{Fig:B_dirty}. The microscopical analysis based on Usadel equation shows that for the perpendicular magnetic configuration $B_y^{\perp}$ component of the induced magnetic field is much more pronounced than $B_z^{\perp}$ which has the same order as $B_z^{\parallel}$ for the parallel configuration (see Fig.\ref{Fig:Q_dirty}). The effect is due to the generation of the long-ranged superconducting correlations in the perpendicular case. We also have demonstrated the enhancement of $B_y^{\perp}$ in comparison with both $B_z^{\perp}$ and $B_z^{\parallel}$ for clean structures (see Fig.\ref{Fig:Q_clean}). In this case the magnitudes of the induced magnetic field oscillate with the period of the order of superconducting coherence length in ferromagnet both in the parallel and perpendicular magnetic configuration. It is interesting to estimate the ratio between the amplitude of these oscillations for the Au/Nb/Co(2$d_0$)/Co($d_0$) structure and parameters relevant to the recent experiment (temperature $T=3$K, superconducting critical temperature $T_c \sim 7.5$K, exchange field in the ferromagnets $h \sim 7.4 \pi T$, superconducting gap $\Delta \sim 1.4 \pi T$), where a significant enhancement ($\sim$ 20 times) of the remote field was observed for the perpendicular magnetic configuration in comparison with the parallel one \cite{Lee_NatPhys}. At $d_0 \sim 0.025 \xi_0$ which corresponds to the maximum values of $B_y^{\perp}$ and $B_z^{\perp}$ the amplitude of the induced magnetic field for the perpendicular case is $\sim 7$ times larger than for the parallel. Thus, our theory provides an adequate explanation of the experimental data in Ref.~\onlinecite{Lee_NatPhys}.
	
	Interestingly, that the direction of the spontaneous magnetic field generated in the superconductor strongly depends on F$_1$ ferromagnet thickness (see Fig.\ref{Fig:Angle}). This effect can be probed experimentally by measuring the critical current $I_c$ of the Josephson junction with an electrode covered by the composite F layer as a function of the external magnetic field ${\bf H}$. As we show, both Fraunhofer dependencies $I_c(H_y)$ and $I_c(H_y)$ are shifted by the values proportional to $Q_y$ and $Q_z$, respectively (see Fig.\ref{Fig:Fraunhofer}). Thus, one can independently determine two components of the spontaneous magnetic field generated in the superconductor due to electromagnetic proximity effect. 
	
	We also demonstrated the possibility of long-ranged superconductivity controlled electromagnetic coupling between two ferromagnets in F$_1$/S/F$_2$ structures. In the case of diamagnetic supercurrents inside the ferromagnets the ground state of the system corresponds to antiferromagnetic ordering (see Fig.\ref{Fig:FSF}). This electromagnetic mechanism of coupling between two ferromagnets should dominate over the exchange one \cite{Zhu} when the thickness of the S layer exceeds $\xi_0$.
	
	 After this work was submitted we learned about a related work Ref.[\onlinecite{Volkov_arxiv}].

	\section*{Acknowledgements}
	
	This work was supported by the French ANR SUPERTRONICS and OPTOFLUXONICS, EU COST CA16218 Nanocohybri, Russian Science
	Foundation under Grant No. 15-12-10020 (S.V.M.), Foundation for the
	advancement of theoretical physics ``BASIS'', Russian Presidential Scholarship SP-3938.2018.5 (S.V.M.), and Russian Foundation for Basic Research under Grant No. 18-02-00390 (A.S.M.). 
	
	\appendix
	
	\section{Solution of the Usadel equation in the dirty limit}\label{App_Dirty}
	
	The anomalous Green function in the $F_1$ layer has the form:
	
	\begin{equation}
	f_s^{(1)}={\rm Re}(F_1), ~~~  f_{ty}^{(1)}=\frac{A_2\sinh(p_1x)}{\sinh(p_1d_1)}, ~~~ f_{tz}^{(1)}=i \rm{Im}(F_1),
	\end{equation}
	while in the F$_2$ ferromagnet the solution of Usadel equation reads as
	\begin{equation}
	\begin{gathered}
	f_s^{(2)}=2\rm {Re}(F_2),\\
	f_{ty}^{(2)}=2i {\rm Im}(F_2)\sin \theta+F_3\cos \theta,\\
	f_{tz}^{(2)}=2i {\rm Im}(F_2)\cos \theta-F_3\sin \theta.
	\end{gathered}
	\end{equation}
	Here we use the following notations:
	\begin{equation}
	\begin{gathered}
	F_1=f_{s0}\cosh (q_1x) +\frac{2A_1\sinh (q_1x)}{\sinh (q_1d_1)},\\
	F_2=\frac{B_1\cosh [q_2(x-d)]}{\cosh(q_2d_2)},\\
	F_3=\frac{B_2\cosh [p_2(x-d)]}{\cosh(p_2d_2)},
	\end{gathered}
	\end{equation}
	where $q_j=\sqrt{2(\omega_n+ih)/D_j}$, $p_j=\sqrt{2\omega_n/D_j}$ ($j=1,2$ is the index corresponding to the F$_1$ or F$_2$ layer), $d=d_1+d_2$, and
	\begin{multline}
	B_1=\frac{f_{s0}}{R(\theta)}\biggl \{Q(p_1,p_2)W(\theta)\cos \theta +\\
	+2 {\rm Re}\left[\frac{Q(q_1,p_2)}{\cosh(q_1^* d_1)} \right]Q(p_1,q_2^*)\sin^2\theta \biggr\},
	\end{multline}
	\begin{equation}
	B_2=-i\frac{2f_{s0}}{R(\theta)} {\rm Im} \left[Q(p_1,q_2)W(\theta) \right]\sin\theta,
	\end{equation}
	\begin{multline}
	A_1=-\frac{f_{s0}}{2}\cosh(q_1d_1)+B_1\cos^2\frac{\theta}{2}+B_1^*\sin^2\frac{\theta}{2} -\frac{B_2}{2}\sin \theta,
	\end{multline}
	
	\begin{equation}
	A_2=2i{\rm Im}(B_1)\sin\theta+B_2\cos\theta,
	\end{equation}
	
	\begin{equation}
	W(\theta)=\frac{2Q(q_1^*,q_2^*)}{\cosh(q_1d_1)}\cos^2\frac{\theta}{2}+\frac{2Q(q_1,q_2^*)}{\cosh(q_1^*d_1)}\sin^2\frac{\theta}{2},
	\end{equation}
	
	\begin{multline}
	R(\theta)=4\biggl[T_1(q_1,q_2)\cos^4\frac{\theta}{2}-T_1(q_1,q_2^*)\sin^4\frac{\theta}{2}\biggr]\cos\theta +\\+4{\rm Re}\biggl[T_2(q_1,q_2)\cos^2\frac{\theta}{2}+T_2(q_1^*,q_2)\sin^2\frac{\theta}{2} \biggr]\sin^2\theta,
	\end{multline}
	
	\begin{equation}
	\begin{gathered}
	Q(\nu_1,\nu_2)=1+\frac{\nu_2 \tanh(\nu_1d_1) \tanh(\nu_2d_2)}{\nu_1},\\
	T_1(\nu_1,\nu_2)=Q(p_1,p_2)|Q(\nu_1,\nu_2)|^2,\\
	T_2(\nu_1,\nu_2)=Q(\nu_1,\nu_2)Q(\nu_1^*,p_2)Q(p_1,\nu_2^*).
	\end{gathered}
	\end{equation}
	
	Next we substitute the expressions for the $\hat f$ into the Eq.(\ref{lambda}) and calculate $Q_1$, $Q_2$ and $Q_3$ using Eq.(\ref{Q_def})
	
	\begin{multline}
	\label{Q1}
	Q_1=\alpha_1 {\rm Re}\sum_{\omega_n>0}  \biggl\{ \frac{f_{s0}^2}{q_1^2 \cosh^2 \chi} \biggl[ q_1^2d_1^2+q_1d_1 \sinh(2q_1d_1+2\chi)-\\-\sinh(q_1d_1+2\chi)\sinh(q_1d_1)\biggr] +\frac{|A_2|^2}{p_1^2\sinh^2(p_1d_1)} \times \\ \times \biggl[p_1^2d_1^2-p_1d_1\sinh(2p_1d_1)+\sinh^2(p_1d_1) \biggr]\biggr\}
	\end{multline}
	
	\begin{multline}
	\label{Q2}
	Q_2=\alpha_2 {\rm Re}\sum_{\omega_n>0} \biggl\{ \frac{4[q_2^2d_2^2+\sinh^2(q_2d_2)]}{q_2^2 \cosh^2(q_2d_2)}B_1^2 -\\-\frac{[p_2^2d_2^2+\sinh^2(p_2d_2)]}{p_2^2 \cosh^2(p_2d_2)}|B_2|^2\biggr\},
	\end{multline}
	
	\begin{multline}
	\label{Q3}
	Q_3=\alpha_2  {\rm Re}\sum_{\omega_n>0} \biggl\{ \frac{4[q_2 d_2-\sinh(q_2d_2)]d_1}{q_2 \cosh^2(q_2d_2)}B_1^2 -\\-\frac{[p_2 d_2-\sinh(p_2d_2)]d_1}{p_2 \cosh^2(p_2d_2)}|B_2|^2\biggr\},
	\end{multline}
	where $\tanh \chi=(2A_1)/[f_{s0}\sinh(q_1d_1)]$, $\alpha_j=(4\pi^2 T \sigma_j)/(c^2) $, $j=1,2$. The values of $Q_y$ and $Q_z$ oscillate as functions of ferromagnets' thicknesses.

	\section{Solution of the Eilenberger equation in the clean limit}\label{App_Clean}
	
	Our goal is to calculate the singlet component of the normal Green function $g_s$ which enters the expression for the supercurrent (\ref{j}). For this purpose we solve the Eilenberger equation in all three layers and find the unknown constants using the boundary conditions on the interfaces.
	
	First, we solve the Eilenberger equation in the superconductor. Since we assume that all Green functions depend only on $x$ and concentrate on the solution in the region with the thickness in the order of $\xi_0 \ll \lambda$ the Eilenberger equations in the S layer reads as
	
	\begin{equation}
	\label{App:S_EEq_g}
	v_x \partial_x g=\Delta^*f-\Delta f^{\dagger},
	\end{equation}
	
	\begin{equation}
	\label{App:S_EEq_f}
	\begin{gathered}
	v_x \partial_x f+2\tilde{\omega }f=2\Delta g, \\
	v_x \partial_x f^{\dagger}-2\tilde{\omega }f^{\dagger}=-2\Delta^* g.
	\end{gathered}
	\end{equation}
	
	Taking the derivative of the Eq. (\ref{App:S_EEq_g}) and substituting $\partial_x f$ and $\partial_x f^{\dagger}$ from Eqs.(\ref{App:S_EEq_f}) we find:
	
	\begin{equation}
	\label{App:S_div_g}
	\partial_{xx}^2 g=\frac{4\Delta^2}{v_x^2}g-\frac{2\tilde{\omega}}{v_x^2}(\Delta^*f+\Delta f^{\dagger}).
	\end{equation}
	
	Performing the same procedure with Eq.(\ref{App:S_div_g}) and using Eq.(\ref{App:S_EEq_g}) we eliminate the anomalous Green functions and obtain the following equation for the normal one:
	
	\begin{equation}
	\label{App:S_CFEq_g}
	\partial^3_{xxx} g-\frac{4\tilde{\Omega}}{v_x^{2}}{\partial_x}g=0,
	\end{equation}
	where $\tilde{\Omega }=\sqrt{\tilde{\omega}^2+\Delta^2}.$  The solution of the above equation which decays at $x \rightarrow -\infty$  reads as
	
	\begin{equation}
	\label{App:S_sol_g}
	g(x)=g_1+g_2e^{kx}
	\end{equation}
	where $k=2\tilde{\Omega}/|v_x|$, $g_1$ and $g_2$ are unknown $2 \times 2$ matrices. Substituting (\ref{App:S_sol_g}) into the Eqs.(\ref{App:S_EEq_f}) we obtain the anomalous Green functions in the superconductor:
	
	\begin{equation}
	\label{App:S_sol_f}
	\begin{gathered}
	f(x)=f_1 e^{-s\kappa x}+\frac{q}{\kappa }g_1+\frac{sq}{k+s\kappa}g_2e^{kx}, \\
	f^{\dagger}(x)=f_2 e^{s\kappa x}+\frac{q^*}{\kappa}g_1-\frac{sq^*}{k-s\kappa}g_2 e^{kx},
	\end{gathered}
	\end{equation}
	where $\kappa=2\tilde{\omega}/|v_x|$ and $q=2\Delta / |v_x|$, $s={\rm sgn}(v_x)$.
	
	With the help of the normalization condition $g^2+f f^{\dagger}=1$ we find that $f_1=f_2=0$,  $g_1 g_2=g_2g_1$ and
	
	\begin{equation}
	\label{App:S_norm_g}
	g_1^2=\frac{k^2}{\kappa^2}\sigma_0.
	\end{equation}
	
	If we represent the unknown matrices as  $g_j=g_{js}+{\bf g}_{jt}\bm{\sigma}$ ($j=1,2$), then from the Eq.(\ref{App:S_norm_g}) with an additional assumption about the singlet structure of the normal Green function deep inside the superconductor we obtain  $g_{1s}=\tilde{\omega} /\tilde{\Omega}$, ${\bf g}_{1t}=0$. Finally, we write down the Green functions (\ref{App:S_sol_g}) and (\ref{App:S_sol_f}) in the usual form $g=g_{s}+{\bf g}_{t}\bm{\sigma}$,   $f=f_{s}+{\bf f}_{t}\bm{\sigma}$,  $f^{\dagger}=\tilde{f}_{s}+\tilde{{\bf f}}_{t}\bm{\sigma}$ and obtain:
	
	\begin{equation}
	\label{App:S_sol_all}
	\begin{gathered}
	g_s=\frac{\tilde{\omega }}{\tilde{\Omega }}+g_{2s}e^{kx}, \\
	\bm{g}_{t}=\bm{g}_{2t}e^{kx}, \\
	f_s=\frac{\Delta }{\tilde{\Omega}}+\frac{sq}{k+s\kappa}g_{2s}e^{kx}, \\
	\bm{f}_{t}=\frac{sq}{k+s\kappa}\bm{g}_{2t}e^{kx}, \\
	\tilde{f}_s=\frac{\Delta^*}{\tilde{\Omega}}-\frac{sq^*}{k-s\kappa}g_{2s}e^{kx}, \\
	\tilde{\bm{f}}_{t}=-\frac{sq^*}{k-s\kappa}\bm{g}_{2t}e^{kx}. \\
	\end{gathered}
	\end{equation}
	where $g_{2s}$ and $g_{2t}$ are the parameters which should be find from the matching conditions.
	
	Now we turn to the solution of the Eilenberger equations in the ferromagnets. The normal Green function in the F layers does not depend on $x$ and equals to $g(0)$. To find the anomalous ones we exclude the vector-potential by gradient transformation. The obtained equations have the standard form. Introducing the following phase:
	
	\begin{equation}
	\beta(a,b)=\frac{e{v_y}}{c v_x}\int\limits_a^b A_{y}(x')dx'+\frac{e{v_z}}{c v_x}\int\limits_a^b A_{z}(x')dx'
	\end{equation}
	we immediately can write down all components of the anomalous Green functions in the F$_1$ layer:
	
	\begin{equation}
	\label{App:F1_sol_f}
	\begin{gathered}
	f_s=(a_1e^{-\varkappa_f x}+a_2e^{-\varkappa_f^* x})e^{-2i\beta(0,x)},\\
	f_{tz}=(a_1e^{-\varkappa_f x}-a_2e^{-\varkappa_f^* x})e^{-2i\beta(0,x)},\\
	f_{ty}=a_3e^{-\varkappa x}e^{-2i\beta(0,x)},\\
	\tilde{f}_s=(b_1e^{\varkappa_f x}+b_2e^{\varkappa_f^* x})e^{2i\beta(0,x)},\\
	\tilde{f}_{tz}=(b_1e^{\varkappa_f x}-b_2e^{\varkappa_f^* x})e^{2i\beta(0,x)},\\
	\tilde{f}_{ty}=b_3e^{\varkappa x}e^{2i\beta(0,x)},
	\end{gathered}
	\end{equation}
	where $\varkappa_f=2(\omega+ih)/v_x$, $\varkappa=2\omega/v_x$. In the F$_2$ layer the anomalous Green functions satisfying the boundary condition in the outer boundary reads as $f_l=F_le^{2i\beta(x,d)}$, $\tilde{f}_l=\tilde{F}_le^{-2i\beta(x,d)}$, where $l=\{s,ty,tz\}$ and
	
	\begin{equation}
	\label{App:F2_sol_f}
	\begin{gathered}
	F_s=m_1e^{-\varkappa_f (x-d)}+m_2e^{-\varkappa_f^* (x-d)},\\
	F_{ty}=[m_1e^{-\varkappa_f (x-d)}-m_2e^{-\varkappa_f^* (x-d)}]\sin \theta +m_3e^{-\varkappa(x-d)} \cos \theta,\\
	F_{tz}=[m_1e^{-\varkappa_f (x-d)}-m_2e^{-\varkappa_f^* (x-d)}]\cos \theta -m_3e^{-\varkappa(x-d)} \sin \theta,\\
	\tilde{F}_s=m_1e^{\varkappa_f (x-d)}+m_2e^{\varkappa_f^* (x-d)},\\
	\tilde{F}_{ty}=[m_1e^{\varkappa_f (x-d)}-m_2e^{\varkappa_f^* (x-d)}]\sin \theta +m_3e^{\varkappa(x-d)} \cos \theta,\\
	\tilde{F}_{tz}=[m_1e^{\varkappa_f (x-d)}-m_2e^{\varkappa_f^* (x-d)}]\cos \theta -m_3e^{\varkappa(x-d)} \sin \theta.
	\end{gathered}
	\end{equation}
	
	Finally, we calculate all unknown constants including desired $g_{2s}$ which enters $g_s$ using continuity of the singlet and the triplet components of the anomalous Green functions in the S/F$_1$ and S/F$_2$ interfaces. The resulting $g_s$ has the form (\ref{gs}).

\end{document}